\documentclass[preprint,12pt,final,5p,times,twocolumn,amsmath,amsfonts,amssymb,fleqn]{elsarticle}

\usepackage{graphicx}
\usepackage{epsfig}
\usepackage{fleqn}
\usepackage{natbib}
\usepackage{txfonts}
\usepackage{geometry}
\usepackage{amssymb}
\usepackage{amsmath}
\usepackage{mathptmx}
\usepackage{amstext}
\usepackage{amsfonts}

\def\beq{\begin{equation}}
\def\eeq{\end{equation}}
\def\bea{\begin{eqnarray}}
\def\eea{\end{eqnarray}}
\journal{Physics Letter A}

\begin{document}

\begin{frontmatter}



\title{Thermodynamic geometry of a kagome Ising model in a magnetic field}


\author{B. Mirza\corref{cor1}}
\ead{b.mirza@cc.iut.ac.ir}
\author{Z. Talaei\corref{}}
\ead{zs$_{-}$talaie@ph.iut.ac.ir}
\cortext[cor1]{Corresponding author}

\address{Department of Physics, Isfahan University of Technology, Isfahan 84156-83111, Iran}

\begin{abstract}
We derived the thermodynamic curvature of the Ising model on a kagome lattice under the presence of an external magnetic field. The curvature was found to have a singularity at the critical point. We focused on the zero field case to derive thermodynamic curvature and its components near the criticality. According to  standard scaling, scalar curvature $R$ behaves as $|\beta-\beta_{c}|^{\alpha-2}$ for $\alpha > 0$ where $\beta$ is the inverse temperature and $\alpha$ is the critical exponent of specific heat. In the model considered here in which $\alpha$ is zero, we found that $R$ behaves as $|\beta-\beta_{c}|^{\alpha-1}$.
\end{abstract}


\end{frontmatter}


\section{Introduction}
Statistical thermodynamics provides the basic tools needed for understanding  macroscopic phenomena in the physical world. Among the geometric representations of  statistical mechanics and thermodynamics of physical systems  known are included  three kinds of Riemannian metric structures to represent fluctuation theorem that have been introduced by Weinhold and Ruppeiner {\cite{weinhold, Ruppeiner}}. One of these metrics  depends on derivatives of internal energy, while the other depends on entropy. Another useful metric was called Fisher-Rao metric \cite{Rao}.  Useful information can be gained about a system by using the thermodynamic curvature $R$ of the space of thermodynamic parameters  expressed through the second and third derivatives of the partition function or an appropriate thermodynamic potential. The space of thermodynamic states is flat if there are no interactions between particles and it curves if interactions are present. The thermodynamic curvature has an interesting behavior close to the critical point which has been shown to be singular for some well-known thermodynamics systems at criticality in which correlation length is infinite (for a useful review see \cite{Ruppeiner1,Brody,Nulton,Rupp}). Furthermore, the thermodynamic curvature determines information about interatomic interaction. Janyszek \cite{Janyszek} found that for an ideal classical gas the scalar curvature $R$ is always zero. Janyszek and Mrugala \cite {Mrugala1} showed that $R$ is negative for Fermi gas and positive for Bose gas. Mirza and Mohammadzadeh \cite{Mirza1,Mirza2} worked out the fractional statistics and also ideal q-deformed bosons and fermions which show the proper sign. It has also been argued that scalar curvature is able to show that the fermion gas is more stable than the boson gas \cite{Mrugala1}. Other thermodynamic models, such as phase transition properties of the van der Waals gas, have been investigated to find that the singular point of the thermodynamic curvature is consistent with the critical point of the system \cite{Brody1,Janke}.\\
A standard scaling for free energy per site $f$ in the critical region may be assumed as
\bea
f(t,h)= \lambda^{-1}f(t\lambda^{a_{t}},h\lambda^{a_{h}})=t^{\frac{1}{a_{t}}}\Psi(h\ t^{\frac{-a_{h}}{a_{t}}})
\eea
where, $t=\beta-\beta_{c}$ and $a_{t}=\frac{1}{\nu d}$, $a_{h}=\frac{\beta\delta}{\nu d}$
are scaling dimensions for energy and spin operators, respectively ($\beta_{c}$ designates the critical point).
$\Psi$ is the scaling function and $d$ is the dimension of the system (more details can be found in \cite{Stanley}. For a second order transition, according to the hypothesis on dimensional grounds, the curvature depends on the correlation volume. Since correlation length $\xi$ is the only physical scale in a system close to the critical region, one can write $R\sim\xi^{d}$. Combined with hyperscaling, $\nu d=2-\alpha$, this leads to
\bea
R\thicksim t^{\alpha-2}
\eea
where, $\alpha $ is the critical exponent related to specific heat. Eq. (2) is confirmed for $\alpha>0$ \cite{Johnston}. It has also been found that this relation is not true for $\alpha<0$ \cite{Mrugala2,Janke1}. In this work, we considered a case in which  $\alpha=0$ to find that the standard scaling would not work for it, either. \\
The 2d Ising model in a magnetic field is a well-known unsolved problem in statistical mechanics. In 1960 Fisher introduced, a remarkable solution for a two-dimensional super-exchange antiferromagnetic Ising model in a magnetic field \cite{Fisher}. The Fisher model is determined in a square lattice where there is an external magnetic field applied to the decorating spins which interact via super-exchange interactions. A similar super-exchange model on the kagome lattice has been considered \cite{Lu,Baxter}. It is known that special cases of the kagome Ising model are soluble in the presence of a magnetic field \cite{Gia,Lin}. In one of the special cases, this model reduces to the Fisher model. In this work, we calculated the scalar curvature for the Fisher model under the presence of an external magnetic field.\\
The rest of the  Letter is organized as follows. In Section 2, the Ising model on the kagome lattice in a nonzero magnetic field is reviewed. In Section 3, the geometrical structure and the thermodynamic curvature are introduced and the scalar curvature is calculated for the Fisher model. The scaling behavior of R is also investigated.
\section{Ising model on the kagome lattice in a nonzero magnetic field }
\begin{figure}[t]
     \center
    \includegraphics[width=0.94\columnwidth]{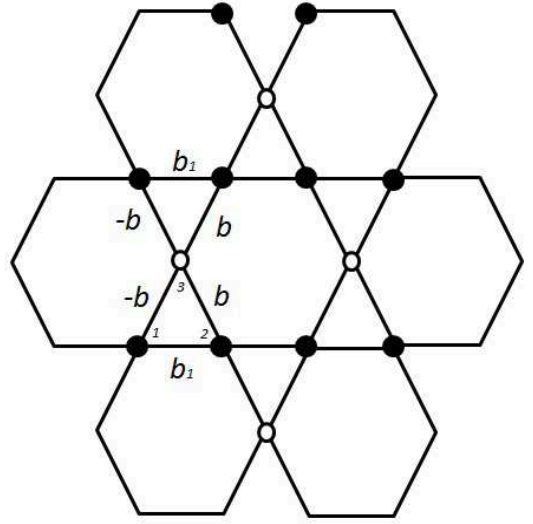}\\
    \caption{The kagome lattice shown in elementary cell with magnetic Ising spins 1,2 (solid circles) and non-magnetic Ising spin 3 (open circles) }\label{figure 1}
   \end{figure}

Consider an Ising model shown in Fig \ref{figure 1} with an interacting energy
\bea
-J_{1}\sigma_{1}\sigma_{2}+J(\sigma_{2}\sigma_{3}-\sigma_{3}\sigma_{1})-\mu H(\sigma_{2}+\sigma_{3})
\eea
around every triangle formed by the spin $\sigma_{1},\sigma_{2},\sigma_{3}$. Assume further that there is an external magnetic field, H, applied to $\frac{2}{3}$ of the lattice sites. The reduced interactions and  field are denoted by $b=\beta J ,b_{1}=\beta J_{1}$ and $h=\beta H$, where $\beta=\frac{1}{kT}$. As a result, the magnetic spins $\sigma_{1}$ and $\sigma_{2}$ interact a super-exchange interaction via the intermediate non-magnetic spin $\sigma_{3}$. $\sigma_{3}$ is an Ising spin which is not coupled to the magnetic field. It has zero magnetic moment. This model does not admit  any restrictions on any positive values of $H$, $J$, $h$, or $b$. The partition function of the kagome Ising lattice has been shown to be equivalent to that of a honeycomb Ising model in a zero field that may be solved as follows:
\bea
Z_{KG} = F^{N} Z_{HC}
\eea
where, $Z_{KG}$ and $Z_{HC}$ are partition functions of the kagome Ising lattice and the honeycomb Ising lattice, respectively, $F^{N}$ is an overall factor, and $N$ is the number of magnetic spins of the kagome lattice or the number of honeycomb lattice sites.\\

The per magnetic spin free energy, $f_{KG}$, and the per site honeycomb lattice free energy, $f_{HC}$, are defined as
\bea
f_{KG} = \lim _{N\rightarrow\infty} N^{-1} \ln Z_{KG}\\
f_{HC} = \lim _{N\rightarrow\infty} N^{-1} \ln Z_{HC}
\eea
Taking $N\rightarrow\infty$ limit of Eq. (4) and making use of the explicit expression of $f_{HC}$ given in \cite{Syozi}, $f_{KG}$ will be
\bea
f_{KG}&=&\ln F + \frac{3}{4} \ln 2 \nonumber\\
&+&\frac{1}{16\pi^{2}} \int_{0}^{2\pi}\ \int_{0}^{2\pi}\ d\theta d\varphi \  \Xi (\theta,\varphi)
\eea
where,
\bea
\Xi (\theta,\varphi)&=&\cosh 2r_{1} \cosh^{2}2r +1 \nonumber\\
&-&\sinh^{2}2r \cos(\theta+\varphi)\nonumber\\
&-&\sinh 2r_{1} \sinh 2r (\cos\theta+\cos\varphi)
\eea
In the above equation, $r_{1}$ and $r$ are functions of $b$ and $b_{1}$. For $b_{1}=0$, this model reduces to the Fisher model for the square lattice. Here, we take the Fisher model and consider the critical behavior of its scalar curvature.
With $b_{1}=0$, the partition function and the free energy per spin of the Fisher model in the square lattice are obtained to be \cite{Lu}
\bea
\label{f} f_{KG}&=&p(b,h)+q(r)
\eea
where
\bea
p(b,h)&=&\frac{3}{2}\ln2\nonumber\\
&+&\frac{1}{4}\ln\{\cosh^2h\ ({\sinh^2h+\cosh^22b})\}\nonumber\\
q(r)&=&\frac{1}{16\pi^2}\int_{0}^{2\pi} \int_{0}^{2\pi}d\theta d\varphi \ln\{\cosh^22r\nonumber\\
&-&\sinh2r\ (\cos\theta+\cos\varphi)\}
\eea
and $r$ satisfies the equation
\bea
e^{-4r}=\frac{\sinh^2h+\cosh^22b}{\cosh^2h}.
\eea
q(r) can be written as
\bea
\frac{1}{16\pi^2}&\times &\int_{0}^{2\pi} \int_{0}^{2\pi}d\varphi d\theta \ln[\cosh^22r\nonumber\\
&\times &\{1-\frac{\sinh2r}{\cosh^22r}(\cos\theta+\cos\varphi)\}]\nonumber\\
=\frac{1}{16\pi^2}&\times &\int_{0}^{2\pi} \int_{0}^{2\pi}d\varphi d\theta \ln\{\cosh^22r\}\nonumber\\
&&\hspace*{-1cm}+\ln\{1-\frac{\sinh2r}{\cosh^22r}(\cos\theta+\cos\varphi)\}\nonumber
\eea
This term can be further simplified to $ \cos\theta+\cos\varphi=2\cos\frac{\theta+\varphi}{2}\cos\frac{\theta-\varphi}{2}$. If we change the variables  $\theta_{1}=\frac{\theta+\varphi}{2}$ , $\theta_{2}=\frac{\theta-\varphi}{2}$ and $\kappa=-\frac{2\sinh2r}{\cosh^22r}$,  we will then have
\bea
q(r)&=&\frac{1}{16\pi^2}\int_{0}^{2\pi}d\theta_{1}\int_{-\pi}^{\pi}d\theta_{2}\{\ln[\cosh^22r]\nonumber\\
&+&\ln[1+\kappa\cos\theta_{1}\cos\theta_{2}]\}\nonumber\\
&=&\frac{1}{2}\ln[\cosh^22r]+\frac{1}{8\pi^2}\int_{0}^{2\pi}d\theta_{1}\int_{-\pi}^{\pi}d\theta_{2}\nonumber\\
&&\ln[1+\kappa\cos\theta_{1}\cos\theta_{2}]\nonumber\\
&=&\frac{1}{2}\ln[\cosh^22r]+\frac{1}{\pi}\int_{0}^{\frac{\pi}{2}}d\theta\nonumber\\
&\times &\ln[\frac{1}{2}(1+\sqrt{1-\kappa^2\sin^2\theta}\ )].\nonumber
\eea
\section{Thermodynamic curvature  }
The geometrical structure of the statistical thermodynamic phase space was investigated in detail by Gibbs  and later developed by Weinhold using energy representation and also by Ruppeiner based on  entropy representation \cite{weinhold,Ruppeiner}.
Alternatively, any  thermodynamic potential representation that is the Legendre transform of entropy or internal energy may also be used. The metric of the representation can be the second derivative of the thermodynamic potential with respect to the  intensive variables. For example, in the Ruppeiner geometry, the extended set of $r+1$ extensive variables of the system can be denoted by $X = (S,N^{1},...,V,...,N^{r})$. The thermodynamic potential is defined as
\bea
\Phi=\Phi({\{F^{i}\}}),
\eea
where, $\Phi$ is a Legendre transform of entropy with respect to the extensive parameter, $X^{i}$, $"r"$ represents the number of various kinds of particles and $F$ is a standard intensive quantity in the entropy representation,
\bea
F^{i}=(1/T,-\mu^{1}/T,...,P/T,...,...\mu^{r}/T)
\eea
\bea
F^{i}=\frac{\partial S}{\partial X^{i}},
\eea
Here, $T$ is  temperature; $\mu$, the chemical potential; and $P$ is  pressure. The $F^{i}$'s are a complete set of coordinates on the thermodynamic phase space. The metric in this representation is given by
\bea
g_{ij}=\frac{\partial^{2}\Phi}{\partial F^{i}\partial F^{j}}.
\eea
Another type of the Riemannian metric in the space of probability distribution is the Fisher-Rao metric. If the distribution has an exponential form such as Boltzmann-Gibbs distribution, the Fisher-Rao metric can be defined as \cite{Rao},
\bea
g_{ij}=\frac{\partial^{2}\ln Z}{\partial \beta^{i}\partial\beta^{j}},
\eea
where, $\beta^{i}=F^{i}/k$ and $Z$ is the partition function. We consider a system with two thermodynamic degrees of freedom. The thermodynamic surface, or the parameter space, is two-dimensional and the metric components are defined by derivative of free energy \bea
f_{ij}\equiv g_{ij}=\partial_{i}\partial_{j}f
\eea
where, $f=\ln Z $  is the reduced free energy per site; and $i$ and $j$ represent $b$ and $h$ which are related to the inverse temperature and the external field, respectively.\\
For such a metric in the two-dimensional thermodynamic space, the Ricci  scalar may be defined by
\bea
R=-\frac{\begin{vmatrix}
                     f_{bb} & f_{bh} & f_{hh} \\
                     f_{bbb} & f_{bbh} & f_{bhh} \\
                     f_{bbh} & f_{bhh} & f_{hhh} \\
                   \end{vmatrix}}{2\begin{vmatrix}
                                    f_{bb} & f_{bh} \\
                                    f_{hb} & f_{hh} \\
                                  \end{vmatrix}^{2}
                   }
\eea
where, $f_{ijk}=\partial_{i}\partial_{j}\partial_{k}f$. To determine $R$, we need to obtain the second and third derivatives of free energy. According to the Eq. (9), the first derivatives of free energy are given by
\bea
&&f_{b}=\frac{\partial f}{\partial b}=\frac{\partial p}{\partial b} + \frac{\partial q}{\partial r}\frac{\partial r}{\partial b}\\
&&f_{h}=\frac{\partial f}{\partial h}=\frac{\partial p}{\partial h} + \frac{\partial q}{\partial r}\frac{\partial r}{\partial h}.
\eea
Other parts of the calculations can be found in the Appendix.
\begin{figure}[t]
     \center
    \includegraphics[width=0.94\columnwidth]{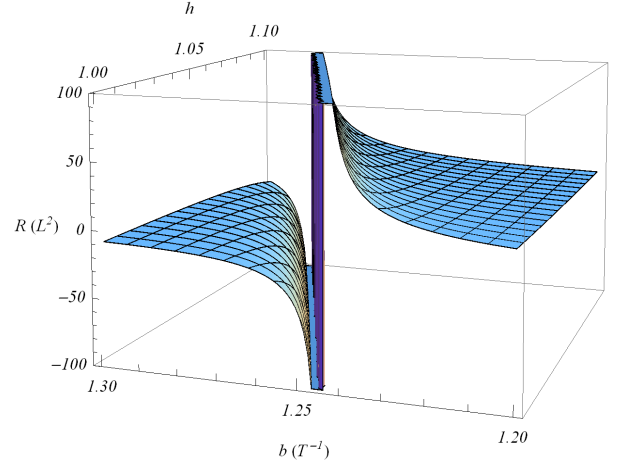}\\
    \caption{A 3d plot of R for $h=1.2$ to 1.3 and $b=1$ to 1.1 .}\label{figure 2}
   \end{figure}
Once the components of $R$ have been determined  for the Fisher model, they are substituted in Eq. (18). The Ricci scalar is depicted as a function of $b$ and $h$ in Fig \ref{figure 2}. From this figure it is found that at high temperature where system is in disordered state (paramagnetic), the sign of $R$ is positive (Fig \ref{figure 4}). As temperature decreases $R$ diverges to plus infinity. At low temperature where system is in the ordered state (antiferromagnetic) $R$ is negative. When temperature increases $R$ diverges to minus infinity. Ordered state is more stable with a lowest ground state energy. Curvature is singular along the critical line in the $h-b$ plane which  exactly corresponds to the Fisher expression \cite{Lu} as shown in Fig.  \ref{figure 3}.\\
\bea
h=\text{arccosh}\ [\sqrt{\frac{\sqrt{2}-1}{2}}\ \text{arcsinh}\ 2b]
\eea
\begin{figure}[t]
    \center
    \includegraphics[width=0.74\columnwidth]{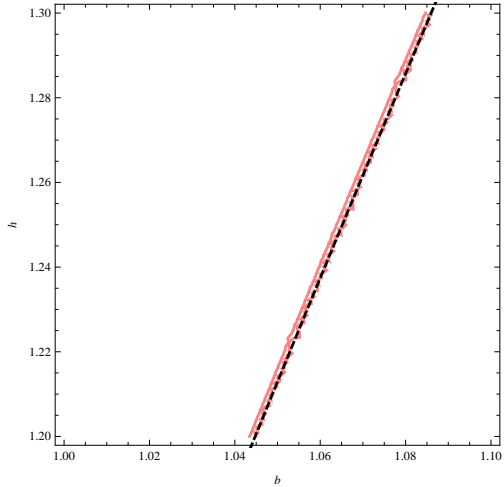}\\
    \caption{A plot of $h$ with respect to $b$ for $R=200$ (gray line) and a plot of $h=\text{arccosh}\ [\sqrt{\frac{\sqrt{2}-1}{2}}\ \text{arcsinh}\ 2b]$ (black dashed line).}\label{figure 3}
   \end{figure}
Focusing  on the zero field case, we have calculated $R$ and its components for this special case. To find the critical point, the denominator of $R$ should be zero and the term that makes it zero is
\bea
&&\cosh^{2}2b+\text{sech}^{2}2b-6 = 0\nonumber\\
&&\Rightarrow ~~~~b_{c}=\text{arccosh}[\sqrt{\frac{2+\sqrt{2}}{2}}\ ]
\eea
where, $b_{c}$ is the critical point that can also be obtained from Eq. (22).
The dominant components of $R$ near the critical point can be written as follows
\bea
f_{bb}\thicksim K(\kappa)\nonumber\\
f_{bh},f_{bbh},f_{hhh}=0\nonumber\\
f_{hh}\thicksim \emph{const}\nonumber\\
f_{bbb}\thicksim \frac{1}{b-b_{c}}\nonumber\\
f_{bhh}\thicksim K (\kappa).
\eea
where $K (\kappa)$ denotes the elliptic integral of first kind. In the above equations, the terms with  odd numbers of $h$ derivatives are equal to zero since there is $\pm h$ symmetry in the free energy and the other terms have expanded about $b_{c}$ to the second order. Then, $R$ is given by
\bea
\hspace*{-0.5cm}R&=\frac{\begin{vmatrix}
             K (\kappa) & 0 & \emph{const} \\
             \frac{1}{b-b_{c}} & 0 & K (\kappa) \\
             0 & K (\kappa) & 0 \\
           \end{vmatrix}
  }{2\ \begin{vmatrix}
                  K (\kappa) & 0 \\
                  0 & \emph{const} \\
                \end{vmatrix}
  ^{2}}.\nonumber\\
\hspace*{-1cm}
\eea
Finally, we  obtained
\bea
R\thicksim \frac{1}{K (\kappa) (b-b_{c})}
\eea
It should be noted that $K (\kappa)$ has a singularity at $\kappa=1$ and can be written as
\bea
K(\kappa)\thickapprox\ln\frac{4}{\sqrt{1-\kappa^{2}}}\thicksim\ln(b-b_{c})
\eea
So, near the critical point, the scalar curvature behaves as in
\bea
R\thicksim\frac{1}{(b-b_{c})\ln(b-b_{c})}
\eea
We can write $\ln x = \lim_{\delta\to0}\frac{1}{\delta}(x^{-\delta}-1)$. So, $\ln (b-b_{c})$ behaves as a power term like $(b-b_{c})^{\delta}$ in which $\delta=0$ \cite{plischke}. We conclude that  $R\thicksim (b-b_{c})^{-1} \equiv t^{-1}$.
\begin{figure}[t]
    \center
    \includegraphics[width=0.74\columnwidth]{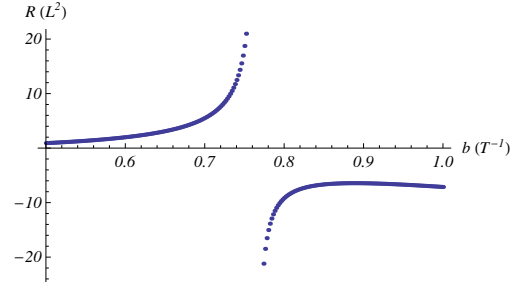}\\
    \caption{A plot of $R$ for $h=0$ and $b=0.5$ to 1 by steps 0.002.}\label{figure 4}
   \end{figure}
For $h=0$, the scalar curvature $R$ is plotted in Fig  \ref{figure 4}. It shows that $R$ diverges at certain point. We  carefully examined this point and found that it coincides with the critical point.
In this work, we  investigated a soluble kagome Ising model under the presence of an external field  in which critical exponent $\alpha$ is zero. There are standard scaling forms for $\alpha > 0 $ and $ \alpha < 0$ as $R \thicksim t^{\alpha-2 }$ and  $R \thicksim t^{\alpha-1 }$, respectively \cite{Johnston,Janke1,Mrugala1}. We found that, for this model with $\alpha = 0$ in 2 dimensions, the scaling behavior of $R\thicksim t^{\alpha-1}$ is similar to $\alpha < 0$ of the spherical model studied in \cite{Janke1}. It should be noted that $\alpha=0$ for the spherical model in $d=4$ and van der Waals model in $d=3$, the scaling behavior of $R$ is $\thicksim t^{\alpha-2 }$ which is the same as $\alpha >0$. However in 2d-Ising model on planar random graph with $\alpha=-1$ and kagome lattice with $\alpha=0$ the scaling behavior is similar as $R \thicksim t^{\alpha-1}$. We think that the scaling behavior of $R$ is also related to dimension. Further studies are needed to explore this behavior.
\vspace{-5 mm}
\section{Conclusion}
The scalar curvature $R$, was derived for the Ising model on the kagome lattice in the nonzero external field. It was found that $R$ has a singularity at critical points (Fig. 2)  and that its singularity corresponds to the Fisher expression  depicted in Fig. 3. For the zero field, we derived $R$ and its components close to the criticality. An expression for $R$ was obtained for this case with $\alpha=0$. This was not what we expected from standard scaling of $R$. It has already been shown that standard scaling does not hold true for negative values of $\alpha$ \cite{Janke1}. Here, we found a similar scaling of the scalar curvature for 2-d Ising model with $\alpha=0$ ($R\sim t^{\alpha-1}$).
\renewcommand{\theequation}{A.\arabic{equation}}
\setcounter{equation}{0}
\section{Appendix A}
 Below are a  few useful relations.
\bea
   &&\frac{\partial q}{\partial b}=\frac{1}{16\pi(\cosh4b+\cosh2h)^{2}}\\
   &&\text{csch}[\frac{1}{2}\log(\cosh^{2}2b\text{sech}^{2}h+\tanh^{2}h)]\nonumber\\
   &&\text{sech}[\frac{1}{2}\log(\cosh^{2}2b\text{sech}^{2}h+\tanh^{2}h)]\nonumber\\
   &&\{\pi(1+\cosh4b+2\cosh2h)^{2}\text{sech}^{2}h-\text{sech}^{2}h-\nonumber\\
   &&16K(4\text{sech}^{2}[\frac{1}{2}\log(\cosh^{2}2b\text{sech}^{2}h+\nonumber\\
   &&\tanh^{2}h)]\tanh^{2}[\frac{1}{2}\log(\cosh^{2}2b\text{sech}^{2}h+\tanh^{2}h)])\nonumber\\
   &&(\cosh4b+\cosh2h-8\cosh^{4}b \text{sech}^{2}h\sinh^{4b})\}\nonumber\\
   &&\sinh4b,\nonumber
\eea
\bea
   &&\frac{\partial q}{\partial h}=\frac{1}{16\pi(\cosh4b+\cosh2h)^{2}}\\
   &&\text{csch}[\frac{1}{2}\log(\cosh^{2}2b\text{sech}^{2}h+\tanh^{2}h)]\nonumber\\
   &&\text{sech}[\frac{1}{2}\log(\cosh^{2}2b\text{sech}^{2}h+\tanh^{2}h)]\nonumber\\
   &&\{\pi(1+\cosh4b+2\cosh2h)^{2}\text{sech}^{2}h-\nonumber\\
   &&16K(4\text{sech}^{2}[\frac{1}{2}\log(\cosh^{2}2b\text{sech}^{2}h\nonumber\\
   &&+\tanh^{2}h)])(\cosh4b+\cosh2h-\nonumber\\
   &&8\cosh^{4}b\text{sech}^{2}h\sinh^{4b})\}\sinh^{2}2b\tanh t,
\eea
\bea
   &&\frac{\partial p}{\partial b} = \frac{\cosh2b\sinh2b}{\cosh^{2}2b+\sinh^{2}h}
\eea
\bea
   &&\frac{\partial p}{\partial h }= \frac{(\cosh^{2}2b+\cosh^{2}h)\tanh h}{\cosh4b+\cosh2h}
   \eea

 \end{document}